\documentclass[letterpaper,english,reprint, aps]{revtex4-2}
\usepackage[T1]{fontenc}
\usepackage[latin9]{inputenc}
\usepackage{color}
\usepackage{amsmath}
\usepackage{amssymb}
\usepackage{graphicx}
\usepackage{wasysym}

\makeatletter

\pdfpageheight\paperheight
\pdfpagewidth\paperwidth

\makeatother

\usepackage{babel}
\begin{document}
\preprint{APS/123-QED}
\title{Detecting subtle macroscopic changes in a finite temperature classical
scalar field with machine learning}
\author{Jiming Yang$^{*}$, Yutong Zheng$^{*}$, Jiahong Zhou}
\thanks{These authors contribute equally.}
\affiliation{Zhixin High School, Guangzhou, Guangdong, China}
\author{Huiyu Li, and Jun Yin}
\affiliation{Atlas Science, Princeton, New Jersey, USA}
\date{\today}
\begin{abstract}
The ability to detect macroscopic changes is important for probing
the behaviors of experimental many-body systems from the classical
to the quantum realm. Although abrupt changes near phase boundaries
can easily be detected, subtle macroscopic changes are much more difficult
to detect as the changes can be obscured by noise. In this study,
as a toy model for detecting subtle macroscopic changes in many-body
systems, we try to differentiate scalar field samples at varying temperatures.
We compare different methods for making such differentiations, from
physics method, statistics method, to AI method. Our finding suggests
that the AI method outperforms both the statistical method and the
physics method in its sensitivity. Our result provides a proof-of-concept
that AI can potentially detect macroscopic changes in many-body systems
that elude physical measures. 
\end{abstract}
\maketitle

\section{Introduction}

Over the past decade, the world has witnessed a remarkable explosion
of advances in artificial intelligence (AI) \citep{lecun2015deep,goodfellow2016deep},
from winning world championships in Go \citep{silver2016mastering,silver2017mastering}
and generating realistic images \citep{sohl2015deep,ho2020denoising,rombach2022high}
to chatbots achieving superhuman performance in many standardized
metrics \citep{vaswani2017attention,bubeck2023sparks}. Accompanying
this AI revolution, a significant amount of research has been dedicated
to applying AI to scientific challenges within the physical sciences
domain, including but not limited to, statistical physics, astrophysics,
particle physics, condensed matter physics, and chemistry \citep{carleo2019machine}.
Within these fields, AI has shown immense potential in recognizing
patterns in complex high-dimensional data and has even contributed
to scientific discoveries \citep{jumper2021highly,wang2023scientific}.

Statistical physics was one of the first fields to intersect with
AI. Even in AI's early years, statistical physics made a mark on its
development \citep{hopfield1982neural,hinton2006reducing,amit1985storing,seung1992statistical,engel2001statistical,gardner1988space}.
In recent years, there has been a renewed intersection of these two
domains. This reunion has seen applications of statistical physics
in studying artificial neural networks\citep{bahri2020statistical}
and using AI to address challenges in statistical physics \citep{carleo2019machine}.
Efforts have been made using AI to detect phase transitions \citep{carrasquilla2017machine,tanaka2017detection,van2017learning}and
to accelerate physical simulations \citep{kochkov2021machine,noe2020machine}.
Specifically, there is a growing interest in leveraging AI techniques
to understand the behavior of many-body systems \citep{carleo2017solving,zhong2021machine,huang2022provably,butler2018machine,zhong2020quantifying,zhong2023non}.
Although machine learning algorithms can easily detect the abrupt
changes in system configurations near phase boundaries, there's a
debate surrounding the necessity of AI in such cases, since traditional
methods measuring order parameters can also detect these changes.
It becomes particularly intriguing to understand up to what limit
can one detect changes in macroscopic quantities, like order parameters,
from observing microscopic configurations in stochastic systems. Identifying
subtle changes in many-body systems proves much more challenging than
pinpointing phase transitions due to microscopic fluctuations sometimes
overshadowing even macroscopic changes. Detecting such subtle macroscopic
changes is of relevance to the study of non-equilibrium many-body
systems, where order parameters are sometimes unknown, and researchers
rely solely on observing the microscopic configurations of the system
\citep{benson2022experimentally,majumdar2018mechanical,paulsen2014multiple,li2022dynamical}.
Examples of these non-equilibrium many-body systems extend from the
classical to the quantum realm, encompassing diverse systems like
soap bubble rafts \citep{mukherji2019strength}, crumpled sheets of
paper \citep{hoffmann2019machine}, DNA self-assembly \citep{zhong2017associative,evans2022pattern},
many-body localized systems \citep{tangpanitanon2020expressibility,zhong2022many},
and quantum many-body scars \citep{serbyn2021quantum}. 

In this paper, we study the sensitivity of physical quantities when
a many-body system undergoes macroscopic changes, such as temperature
variations. We adopt a (classical) 2-dimensional scalar field at finite
temperature discretized on a lattice as a coarse-grained toy example
of a many-body system, examining its configuration alterations under
varied temperatures. We focus on the two-point correlation function
as an example of physical quantity \citep{sethna2021statistical},
using its values from different samples to identify temperature changes
in the scalar field. We also propose using an autoencoder's latent
space to detect temperature changes in scalar field samples, and compare
its accuracy with the two-point correlation function. Our result suggests
that while the two-point function can identify subtle temperature
variations, our AI method offers higher sensitivity, outperforming
traditional physical measurements in detecting subtle macroscopic
changes in the scalar field. Our result provides a proof-of-concept
that AI can potentially detect macroscopic changes in many-body systems
that elude physical measures. Generative models of scalar fields has
also been studied in the context of lattice field theories \citep{zhou2019regressive,komijani2023generative,pawlowski2020reducing,del2021efficient},
but with a different focus than the present work. 

The organization of the paper is as follows: In Section \ref{subsec:Massive-scalar-field}, we present the details of our lattice simulation of the classical $\phi^{4}$ scalar field in 2D and discuss how to compute the two-point correlation function numerically for the scalar field. In Section \ref{subsec:Problem-setup}, we set up the problem we are trying to study. In Section \ref{subsec:Autoencoder}, we present details of our autoencoder and discuss its latent representations. Then, in Section \ref{sec:Binary-classification}, we discuss how to use the latent space, correlation function, and Principal Component Analysis (PCA) for classification of sampled configurations from the scalar field at different temperatures. Lastly, in Section \ref{subsec:Correlation-function-scaling}, we discuss the finite-size effects of the two-point correlation function.

\section{Methods}

\subsection{$\phi^{4}$ scalar field in 2D}

\label{subsec:Massive-scalar-field}

The scalar field is widely used in various areas of physics, applicable to describe diverse phenomena ranging from temperature and pressure, electric potential in electrostatics, and Newtonian gravitational potential, to the Standard Model in particle physics \citep{srednicki2007quantum,kardar2007statistical}. In our study, we primarily focus on the classical scalar field in the context of statistical mechanics. Let us consider a scalar field with $\phi^{4}$ interaction in contact with a heat bath at finite temperature $T$. The Hamiltonian is given by 
\begin{equation}
H=\frac{1}{2}m^{2}\phi^{2}+\frac{1}{2}\left(\partial_{\mu}\phi\right)^{2}+\text{\ensuremath{\lambda\phi^{4}}},\label{eq:Hamilton}
\end{equation}
where $m$ represents the mass of the scalar $\phi$ and $\lambda$
is the interaction strength. From the perspective of statistical mechanics,
the mass term $m^{2}\phi^{2}/2$ describes tendency of the scalars
to achieve different values, and the the kinetic term $\left(\partial_{\mu}\phi\right)^{2}/2$
describes the cooperativity among the scalars to achieve the same
value.

\subsubsection{Two-point correlation function}

To detect macroscopic changes in scalar configurations, it is useful to consider the two-point correlation function, which measures the level of correlation among scalars at a fixed distance. For instance, in the case of configurations of ferro- and antiferromagnetic materials, it helps to determine whether the spins are more likely to align with or repel their neighbors. Consider the value of the scalar field at position $x$ and another position distance $d$ away from it; then the two-point correlation function is given by

\begin{equation}
C_{d}:=\left\langle \phi(x)\phi(x+d)\right\rangle ,\label{eq:Correlation_Func}
\end{equation}
where $\left\langle \cdot\right\rangle $ refers to averaging over
different thermal realizations. In our study, we discretize the space
to a 2D lattice, as our system is translationally-invariantand, we
also perform average over spatial location $x$ and lattice sites
distance $d$ away from $x$. To normalize our measurements, we divide
the results by $C_{0}=\left\langle \phi^{2}(x)\right\rangle $, which
is the self-correlation. At high temperatures, the system is dominated by thermal fluctuations and, consequently, tends to become more disordered. Therefore, at high temperatures, there is less correlation between the scalars, resulting in a smaller value for the two-point correlation function. In the extreme case, as the temperature approaches infinity $(T\to\infty)$, the $C_{d}/C_{0}$ approaches zero. On the
other hand, as the temperature approaches zero $(T\to0)$, thermal
fluctuation is absent and the scalars is domniated by the interactions
in Eq.\ref{eq:Hamilton}, $C_{d}/C_{0}$ approaches a finite value.
This suggests an extremely high level of cooperativity among the scalars.

\subsection{Problem setup}

\label{subsec:Problem-setup}

Between the temperature range $(0,\infty)$, there is a phase transition from a low-temperature phase to a high-temperature phase for the scalar field. This transition illustrates a shift from the most ordered state to the most disordered state and can be observed by analyzing the values of the two-point correlation function Eq.\ref{eq:Correlation_Func}. However, differentiating subtle macroscopic changes in the same phase is more difficult than detecting different phases in the presence of noise. Therefore, a question arises: can we determine, based on samples of the scalar field, whether these samples were prepared at the same or different temperatures? It is relatively easy for experimenters to distinguish samples from different phases, either through visual inspection or by measuring order parameters. However, distinguishing samples prepared at nearby temperatures is challenging, as the correlation function values will also tend to be close. In such situations, we propose the use of AI tools to make such judgments based on scalar field samples prepared at the two temperatures. In this study, we compare three families of methods for detecting temperature changes in the scalar field: AI methods such as autoencoders (see in Section \ref{subsec:Autoencoder} and Appendix), traditional statistical methods such as t-distributed Stochastic Neighbor Embedding (t-SNE, see Section \ref{subsec:t-SNE} and Appendix \ref{subsec:t-SNE-embedding Introduction}), and Principal Component Analysis (see Appendix \ref{subsec:PCA-and-correlation Introduction}), and physical measurements such as the two-point correlation function.

The problem setup is as follows: We prepare two sets of scalar field samples from different temperatures, and the task is to determine whether two randomly drawn samples are from the same or different temperatures, and to quantify the rate of successful differentiation.

To build up the dataset, we perform Monte-Carlo simulation (see Appendix
\ref{subsec:Monte-Carlo-simulation Introduction}) of the scalar field
on a 2D lattice (lattice size $N=28\times28$), for twenty different
temperature points chosen unifromly on a log scale from $10^{-4}$
to 1. Then, we fix one set of the data $\mathcal{D}_{h}$ as the configuration
of the highest temperature $T_{h}=1$, and another set of data $\mathcal{D}_{l}$
as configurations of a lower temperature $T_{l}$ chosen from the
remaining temperatures. Then, for a pair of samples randomly drawn
from $\mathcal{D}_{h}\cup\mathcal{D}_{l}$, our task is to determine
whether they are from the same or different temperatures ($T_{h}$
or $T_{l}$).

\begin{figure}[b]
\includegraphics[scale=0.17]{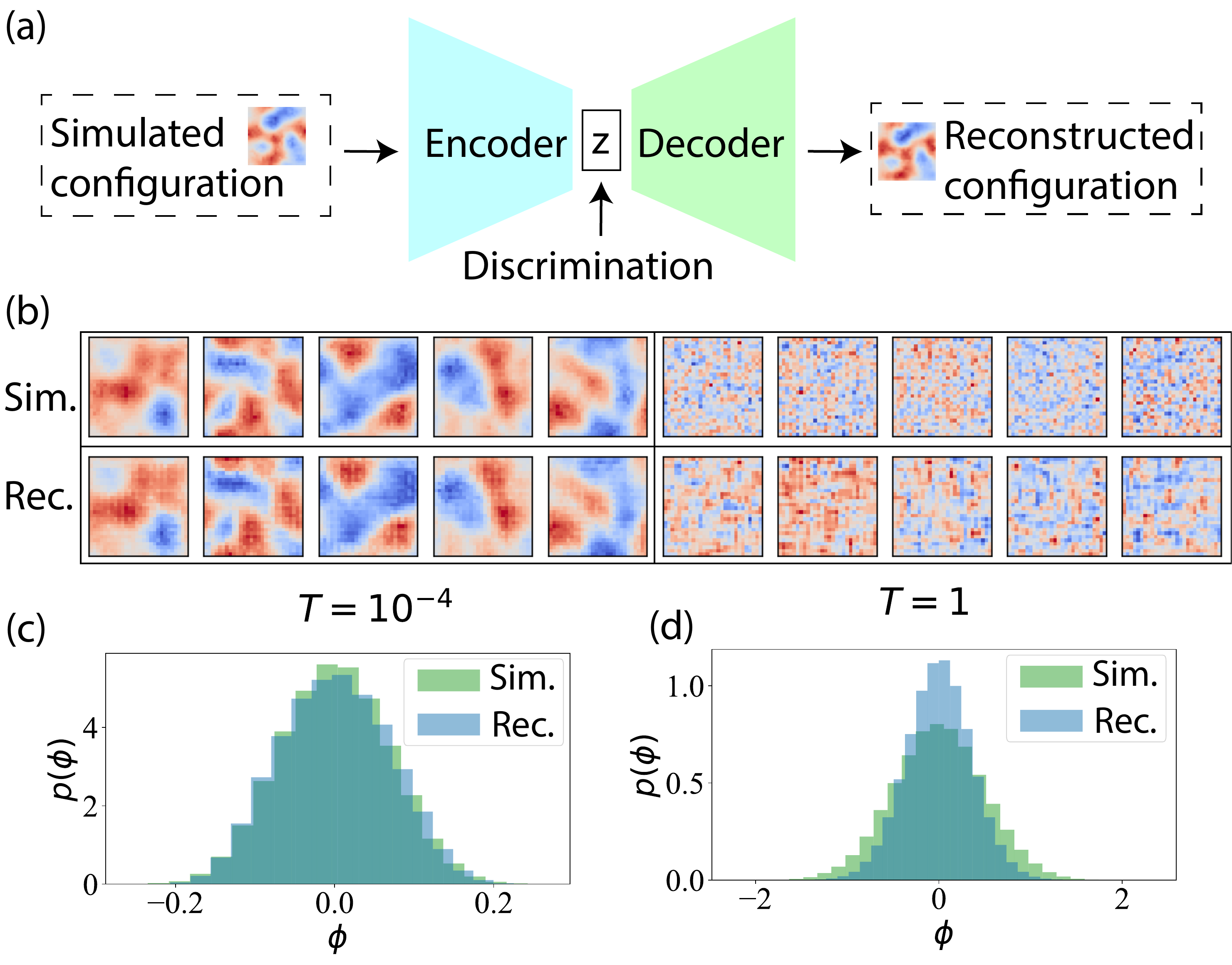}

\caption{\label{fig:autoencoder} Learning scalar field configuration with
autoencoder. (a) schematics of the autoencoder. (b) reconstructed
scalar field configuration at different temperatures. (c)-(d) the
distribution of simulated vs reconstructed scalar field configurations
at (c) $T=10^{-4}$ and (d) $T=1$.}
\end{figure}

\subsection{Autoencoder}

\label{subsec:Autoencoder}

The autoencoder is an artificial neural network that is trained to reconstruct its input data in order to learn a compressed representation of the data \citep{hinton2006reducing,goodfellow2016deep,bank2023autoencoders}. It offers a more flexible and computationally efficient approach to dimensionality reduction. We treat the scalar field configurations as pixel intensities and task the autoencoder with reconstructing the configuration of our $\phi^{4}$  scalar field (see schematics in Fig.\ref{fig:autoencoder} (a)). We use convolutional layers in the autoencoder (see Appendix \ref{subsec:Convolutional-Autoencoder introduction}), which promise better reconstruction quality compared to fully-connected layers.

Before attempting to answer the classification question in Sec.\ref{subsec:Problem-setup},
we first show that our autoencoder is capable of capturing the statistics
of the scalar field configurations. Fig.\ref{fig:autoencoder} (b)
shows the learning outcomes at different temperatures (5 pairs of
random examples at each temperature: top row is simulation, bottom
row is reconstruction from autoencoder). The temperature on the right
five subfigures is high ($T=1$), and the temperature on the left
five subfigures is relatively low ($T=10^{-4}$). For $T=10^{-4}$,
the reconstruction quality is high, and the reconstructed configurations
almost perfectly resemble the simulated ones. However, for $T=1$
there are still noticeable discrepancies from the simulation, and
a possible cause for this discrepancy might be due to the fact that
convolutional layers tend to introduce more correlation into the configuration
than is present in the simulation in this high temperature phase.

In Fig.\ref{fig:autoencoder} (c)-(d), we show the distribution of
all 300 pairs of simulated and reconstructed configurations at $T=10^{-4}$
and $T=1$. We can see that for $T=10^{-4}$, the two distributions
match fairly well, while at $T=1$ there are stronger discrepancies,
which agree with our observation in Fig.\ref{fig:autoencoder} (b).

\subsubsection{t-SNE embedding}

\label{subsec:t-SNE}

The latent space in the convolutional autoencoder captures the most salient features of the data in low dimensions. In the convolutional autoencoder described above, although the latent space has only two neurons, the convolutional layers still make the latent space $2\times7\times7=98$
dimensional and therefore difficult to interpret by humans. In the following, we use t-SNE embedding \citep{hinton2002stochastic} to perform dimensionality reduction of the latent space into 2-dimensional (see Appendix \ref{subsec:t-SNE-embedding Introduction}). Each input configuration becomes a single point in the two-dimensional t-SNE embedded latent space. The data from high-temperature configurations
$\mathcal{D}_{h}$ and low-temperature configurations $\mathcal{D}_{l}$
are observed as two distinct groups of points in the embedded space. The goal is to classify these points based on their latent space positions. To achieve this, we can train a binary classifier using the logistic regression algorithm \citep{menard2002applied} (see Appendix \ref{sec:Binary-classification}). After the completion of training, the classifier is capable of discerning whether a pair of novel configurations belong to identical or distinct temperatures through examining their positions in the latent space. In the following, we detail this approach which allows for efficient and effective binary classification based on the latent representations of the input configurations.

\subsection{Novelty detection: is the temperature same or different?}

Logistic regression is a binary classification algorithm that uses
a linear decision surface to distinguish two classes of input patterns.
Specifically, we treat the latent space points \textbf{$\boldsymbol{z}^{\mu}$}
($\mu=1,...,300$) as the input for binary classification (see Appendix.\ref{subsec:Binary-classification Introduction}).
Given data in $\mathcal{D}_{h}$ and $\mathcal{D}_{l}$, we assign
the latent representations for samples in $\mathcal{D}_{h}$ with
the label $y^{\mu}=+1$ (after passing $\boldsymbol{x}^{\mu}$ from
$D_{h}$ through the autoencoder, we obtain $z^{\mu}$ in the latent
space and assign $y^{\mu}=1$), and for $\mathcal{D}_{l}$ with the
label $y^{\mu}=-1$. For these two classes of $z^{\mu}$, we perform
logistic regression to classify them as shown in (Fig.\ref{Figure:latent_space_panel}),
and determine the number of correctly classified points. Based on
the percentage of correct classification, we can assign a score (see
Fig.\ref{Figure:latent_space_panel}).

\subsubsection{Comparison with statistical methods}

Is the AI method really necessary? We compare our AI method with directly using traditional statistical methods like Principal Component Analysis (PCA) \citep{shlens2014tutorial} and t-SNE on the raw data, and show that the AI method indeed outperforms such traditional methods. After dimensionality reduction by PCA or t-SNE on the raw data, we perform binary classification as above. Based on the percentage of correct classifications, we can assign scores to PCA (red line in Fig.\ref{Figure:latent_space_panel}(a)) and t-SNE (purple line in Fig.\ref{Figure:latent_space_panel}(a)).

\subsubsection{Novelty detection with correlation function value}

\begin{figure}
\includegraphics[scale=0.23]{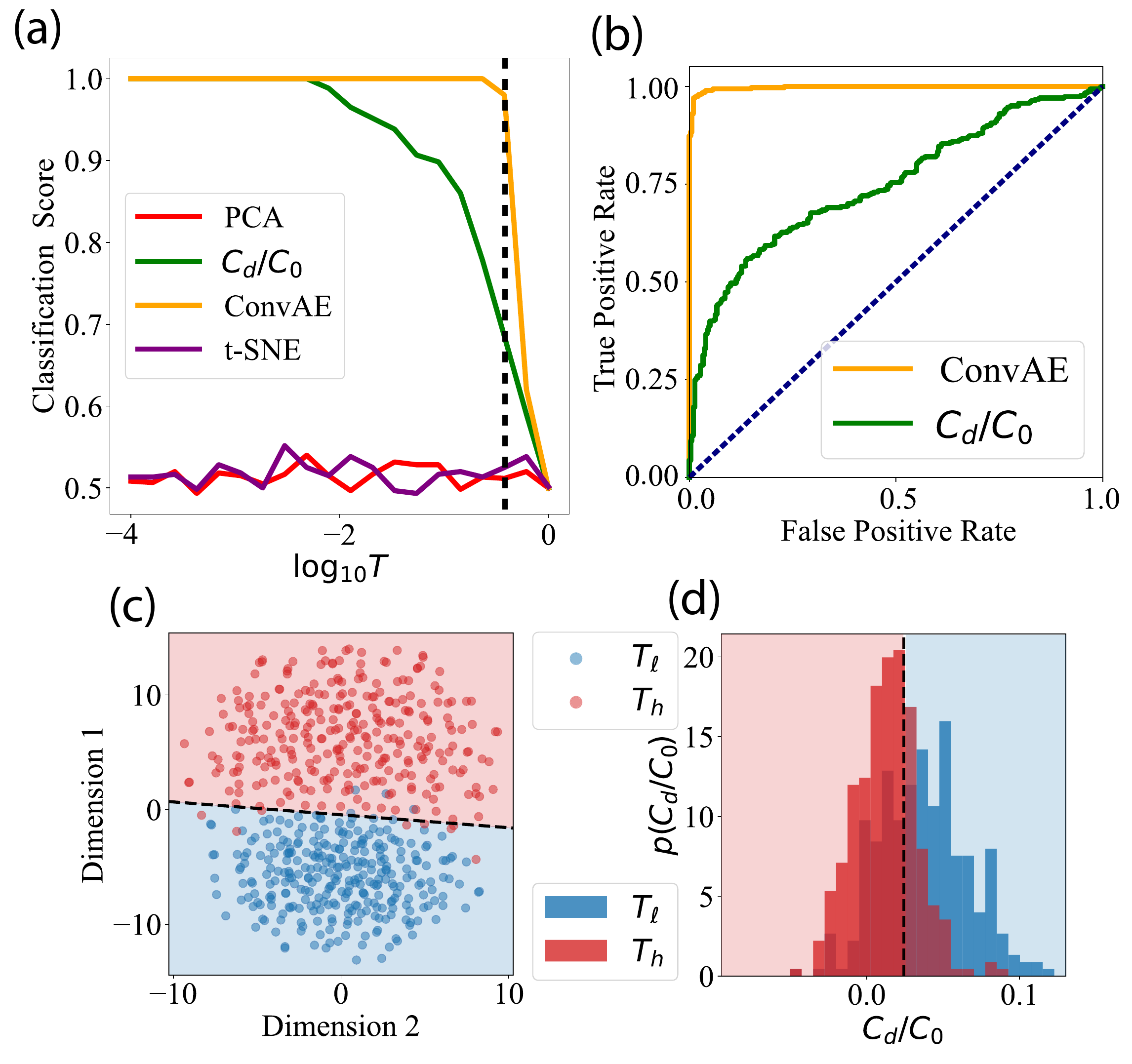}\caption{\label{Figure:latent_space_panel} Differentiating temperatures of
scalar field sample using different methods. (a) The accuracy score
of t-SNE, convolutional autoencoder, PCA, and correlation function
as a function of $\mathcal{D}_{l}$'s temperature $T_{l}$.\textcolor{red}{{}
}(b)-(d): Example ROC curve (b), t-SNE embedding of the autoencoder
latent space (c), and distribution of $C_{d}/C_{0}$ at $T_{l}=0.40$ (corresponds
to the dashed line in (a)) (b) The ROC curve of fully-connected autoencoder
(yellow line) and $C_{d}/C_{0}$ (green line), dashed line shows the result
of a random classifier. (c) The t-SNE embedding of the autoencoder
latent space in Section \ref{subsec:t-SNE}. There are two cluster
of points with temperature $T_{h}=1$ (red points) and $T_{l}=0.40$
(blue points), and we can see that the points from different datas
have little overlap among the decision surface, indicating a good
classification result.\textcolor{red}{{} }(d) The distribution of $C_{d}/C_{0}$
values. Red histograms ($T=1$) corresponds to measurements of data
from $D_{h}$ and blue histograms ($T=0.40$) corresponds to measurements
of datas from $D_{l}$.\textcolor{red}{{} }There is significant overlap
between this two clusters, representing a not ideal classification
result.}
\end{figure}

Additionally, for every scalar field configuration in $D_{h}$ and
$D_{l}$, we can measure its $C_{d}/C_{0}$ value (Eq.\ref{eq:Correlation_Func},
we focus on $d=4$). Now we have two different groups of $C_{d}/C_{0}$
values which form two different histograms in the distribution of
$C_{d}/C_{0}$ (Fig.\ref{Figure:latent_space_panel}(d)). Then we can perform
binary classification on these values and classify which $C_{d}/C_{0}$
values belong to which dataset ($D_{h}$ or $D_{l}$). The score of
$C_{d}/C_{0}$ is shown in the green line in Fig.\ref{Figure:latent_space_panel}(a).
Additionally, we report the Receiver-Operator-Characteristic (ROC)
curve (see Appendix \ref{subsec:ROC-curve Introduction}) for the
novelty detection result of the convolutional autoencoder and the
correlation function (Fig.\ref{Figure:latent_space_panel}(b)).

Fig.\ref{Figure:latent_space_panel}(a) shows three curves representing
accuracy scores, which indicate the percentage of correct classifications
for distinguishing data from $\mathcal{D}_{h}$ or $\mathcal{D}_{l}$
of scalar field configurations. These classifications are obtained
using the autoencoder method (the AI method), the principle component
analysis method (the statistical method), and the two-point correlation
function method (the physics method) under different $T_{l}$ values.
In this case, the high-temperature sample ($\mathcal{D}_{h}$) is
always set at a temperature of $T_{h}=10^{0}$. We take 20 uniformly
distributed points in log space from $[10^{-4},10^{0}]$ and set each
of them as the $T_{l}$ for comparison.

The scores $1$ and $0.5$ are two special reference points. A score
of $1$ means that our classification is $100\%$ correct, while a
score of $0.5$ means that we are guessing blindly and the correctness
is equivalent to chance ($50\%$ for binary classification). For the
green and yellow lines, there is a relatively stable score within
a certain temperature range, followed by a sharp decrease to score
$0.5$ as the temperature approaches a certain point for both methods.
However, the reduction occurs earlier for the two-point correlation
method (around $T=10^{-2}$) compared to the convolutional autoencoder
method (around $T=10^{-1.8}$). Additionally, the green line decreases
sharply within the temperature range of $[10^{-2},10^{0}]$, and the
yellow line falls similarly only when $T\apprge10^{-0.4}=0.6$. Generally,
there is a trend of declining correctness as the temperature increases.
Both the PCA method and t-SNE method shows a relatively uniform fluctuation
in the score, centered around $0.5$ and oscillating within a range
of no more than $0.15$. This is much lower compared to the scores
of the convolutional autoencoder method and the two-point correlation
function method. We only utilize the first and second principal components
for the PCA and t-SNE classification for visualization purposes. However,
as shown in the figure, the score is only $0.5$, indicating that
two dimensions are not sufficient for these methods.

To show the effectiveness of the convolutional autoencoder, we fix
$T_{l}=10^{-0.4}$ to make comparison of the methods' performance,
as shown in Fig.\ref{Figure:latent_space_panel}(b)-(d). In Fig.\ref{Figure:latent_space_panel}(b), there is a ROC curve (see Appendix
\ref{subsec:ROC-curve Introduction}) evaluating the performances
of two-point correlation function method and convolutional autoencoder
under the temperature such that $T_{h}=10^{0}$ and $T_{l}=10^{-0.4}$.
Here, the diagonal (in dashed line) between the coordinate axises
refers to the rate of correct classification when making blinding
guess, and the ROC curves of convolutional autoencoder and that of
two-point correlation function method are shown in yellow and green,
respectively. In the diagram, the Area Under the Curve (AUC, also
see Appendix.\ref{subsec:ROC-curve Introduction}) for the curve of
convolutional autoencoder is 0.997, which is significantly higher
than the AUC for the curve of two-point correlation function, which
is 0.748. This gives strong support to our conclusion that convolutional
autoencoder is a much better classifier than correlation function
in this context.

In Fig.\ref{Figure:latent_space_panel}(c), the points line above
the dashed line of the plot are all data from $D_{h}$ (in red) with
$T=1,$ and below are from $D_{l}$ with $T=10^{-0.6}$ (in blue).
The dotted line represents logistic regression performed on these
two classes of points. As the temperature increases, the red points
and blue points in the latent space get closer. If the temperature
continues to increase, those points will eventually merge together.
This is also why, as the temperature increases, it becomes more difficult
for the convolutional autoencoder to correctly classify, and the scores
decrease sharply.

In Fig.\ref{Figure:latent_space_panel}(d), the histograms show the
distribution of values of the correlation function for samples in
$D_{h}$ (in red) with $T=1,$ and $D_{l}$ with $T=10^{-0.6}$ (in
blue). There is also a decision surface in between the two histograms
to separate the $C_{d}/C_{0}$ values from the $D_{h}$ and $D_{l}$ data.
The points to the left of the decision surface belong to $D_{h}$
and the points to the right belong to $D_{l}$. As we can see, there
is a significant overlap between the two histograms, and the classification
fails in this overlap region, resulting in a lower classification
score.

The PCA score shows a relatively uniform fluctuation around 0.5, which is essentially equal to chance, indicating that the first two principal components alone are insufficient for correct classification. Since a linear autoencoder is equivalent to PCA, when comparing the classification performance of the two-dimensional embedded latent space of the (nonlinear) autoencoder, we conclude that nonlinearity is necessary for this task. Thus, we cannot use traditional methods like PCA for this classification task while requiring good visualization of the decision process (2-dimensional). Similarly, performing t-SNE directly on the original data yields poor classification results. This justifies our use of AI methods (autoencoder) over traditional statistical methods (PCA and t-SNE). In conclusion, we find that autoencoder-based classification of different temperatures outperforms both statistics methods (PCA and t-SNE) and physical metrics (correlation function).

\section{Conclusion and discussion}

\subsection{Conclusion}

In this paper, we study the problem of differentiating between different temperatures in scalar field samples, serving as a toy model for detecting subtle macroscopic changes in a many-body system. We used a convolutional autoencoder (AI method), PCA and t-SNE (statistical methods), and the two-point correlation function (physics method) to distinguish samples of scalar field configurations with increasingly close temperatures. Additionally, we also attempted to compress the high-dimensional configuration of the scalar field and reconstruct it from the low-dimensional latent space of an autoencoder. Based on the results we have collected, we show that the autoencoder is able to capture realistic statistics of the scalar field, similar to physical simulations. We also demonstrate that both the latent space of the autoencoder and the two-point correlation function can be used to perform classification of different temperatures for the scalar field, while the AI method (convolutional autoencoder) outperforms the physical method (two-point correlation) as well as the statistical methods (PCA and t-SNE).

Our study can still be further optimized in a number of ways. The statistics of the high-temperature phase of the scalar field system can potentially be better captured by improving the structure of the autoencoder. Furthermore, while we focus only on a scalar field at equilibrium, we can couple the scalar field to an external source to drive the system out of equilibrium, which would be of interest for more realistic non-equilibrium many-body systems. We can also extend our method to more sophisticated models, such as vector fields (multiple component fields), and use the AI method we presented to differentiate other physical quantities such as magnetization, free energy, and dissipation. Finally, it would also be interesting to explore whether our results can be applied to detect macroscopic changes in experimental many-body systems.

\begin{acknowledgments}
The authors acknowledges support from Zhixin High School Science Outreach
Program. J.Y., Y.Z, and J.Z. would like to thank Weishun Zhong for
suggesting this project and guidance throughout carrying out the simulation
and preparing this manuscript. 
\end{acknowledgments}

\appendix

\section{Correlation function scaling}

\subsection{Scalar field simulation}

\subsubsection{Monte-Carlo simulation\label{subsec:Monte-Carlo-simulation Introduction}}

To obtain the massive scalar field system in equilibrium, we utilize
the Metropolis-Hastings algorithm. This algorithm employs Monte Carlo
methods to update the configuration of these scalars $\phi$ with
random samples obtained from the Boltzmann distribution $P$ determined
by the Hamiltonian $H$ of the field

\begin{equation}
P(\phi)=e^{-\beta H(\phi)}.\label{eq:Prob_dstrib_Monte_Carlo}
\end{equation}

Here, $\beta$ stands for the inverse temperature defined as $1/K_{B}T$,
where $K_{B}$ refers to Boltzmann constant, and $T$ refers to temperature
of the system.

To start our simulation, we initialize a random configuration of scalars
$\phi$ that follows a normal distribution. We then calculate the
Hamiltonian of this configuration, denoted as $H[\phi]$ (Eq.\ref{eq:Hamilton}),
along with its associated probability $P(\phi)$ (Eq.\ref{eq:Prob_dstrib_Monte_Carlo}).
During each Monte-Carlo sweep, we randomly propose a new configuration
$\phi'$ drawn from normal distribution, and calculate the Hamiltonian
of the current configuration $H[\phi']$ and its associated probability
$P(\phi')$.

If $\triangle H<0$, the new configuration $\phi'$ is always accepted
and used as the new $\phi$ for the next sweep. 

If $\triangle H>0$, by contrast, our acceptance is determined by
$r$, the acceptance probability is calculated as

\begin{equation}
r=\frac{P(\phi')}{P(\phi)}=e^{\beta(H[\phi']-H[\phi])}\equiv e^{\beta(\triangle H)},\label{eq:Generated_Random_Number_Monte_Carlo}
\end{equation}

where $\triangle H=H[\phi']-H[\phi]$. 

Throughout all the simulations conducted in this paper, we choose
$m^{2}=0.1$ and $\lambda=0.01$. For each temperature ($20$ of them)
used in the current study, we perform $300$ independent realizations
of the scalar field simulation.

\label{subsec:Correlation-function-scaling}

\subsubsection{Finite size effects}

\begin{figure}
\includegraphics[scale=0.2]{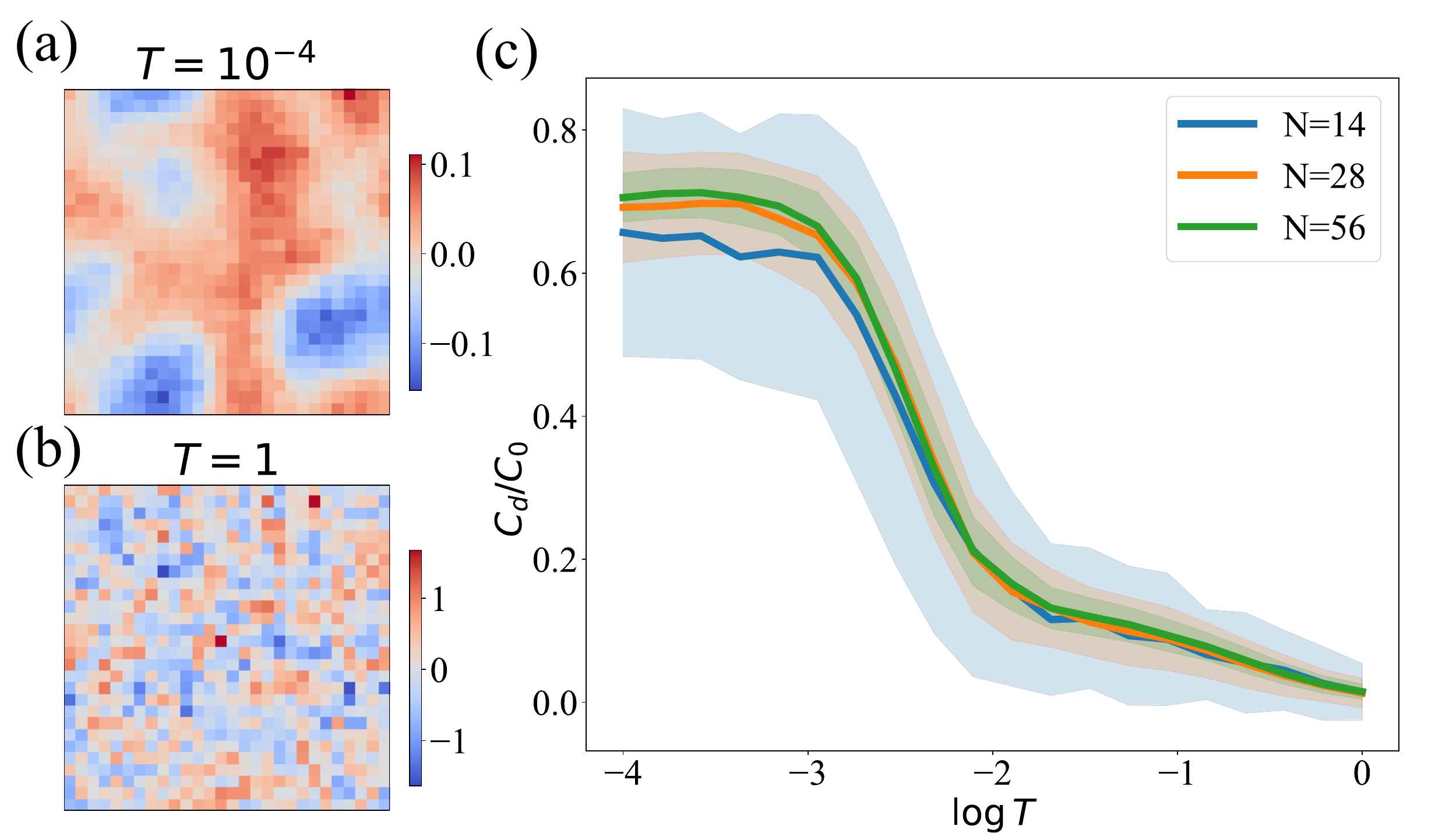}\caption{\label{Figure: scalar_field_panel} Differentiating sample temperatures
among different methods. (a) typical configuration of $\phi$ in a
low-temperature($T=10^{-4}$) phase. (b) typical configuration of
$\phi$ in a high-temperature ($T=10^{0}$) phase. (c) Average value
of two-point correlation function for $\phi$ in different temperatures($T\in[10^{-4},10^{0}]$)
for three different sizes of the system($N=14,28,56$), one standard
deviation is included as shade.}
\end{figure}

In this section, we study potential finite size effects on the two-point
correlation function (Eq.\ref{eq:Correlation_Func}). We simulate
the scalar field in Eq.\ref{eq:Hamilton} for different lattice sizes
and report the measured $C_{d}/C_{0}$ in Fig.\ref{Figure: scalar_field_panel}.

In Fig.\ref{Figure: scalar_field_panel} (a)-(b), we show the configurations
of the highest and lowest temperatures of the scalar field we studied.
When comparing Fig.\ref{Figure: scalar_field_panel}(a) with Fig.\ref{Figure: scalar_field_panel}(b)
, it is evident that the phase shown in (a) is more ordered than (b).
There are many large areas with similar colors, indicating similar
scalar values between neighboring scalars in these areas. This is
because at low temperature ($T=10^{-4}$), there are negligible random
noises compared to the system at high temperature $(T=10^{0})$, and
cooperativity dominates the interactions between scalars. Additionally,
a much smaller range of $\phi$ values is observed in the lower temperature
phase. In contrast, the pattern in Fig.\ref{Figure: scalar_field_panel}
(b) appears to be very chaotic, with the values of most neighboring
scalars varying much more, leading to fewer coherent areas.

In Fig.\ref{Figure: scalar_field_panel}(c), we show values of the
two-point correlation function (Eq.\ref{eq:Correlation_Func}) for
$N=14,28,56$ when the temperature changes in the range of $[10^{-4},10^{0}]$.
Solid lines are the mean value and shades correspond to one standard
deviation. Initially (at around $T=10^{-4}$), the mean correlation
value is higher for larger sizes. However, although the value still
increases with increasing size at the same temperature, the difference
between the three lines in the vertical section becomes negligible
as $T$ increases. Furthermore, the phase-transition temperatures
for these three conditions are similar to the others, occurring at
approximately $T=10^{-2.5}$. Therefore, we conclude that within the
purpose of this study, the effect of different sizes $N$ can be neglected
when using $C_{d}/C_{0}$ for differentiating different scalar field temperatures.
Thus, we fixed $N=28$ for all other simulations performed in this
paper.

\subsection{Autoencoder\label{subsec:Autoencoder Introduction}}

\label{subsec:Autoencoder_app}

The autoencoder consists of an encoder $g(\boldsymbol{x}^{\mu})=\boldsymbol{z}^{\mu}$,
which compresses the input data $\boldsymbol{x}^{\mu}$ into a latent
feature representation $\boldsymbol{z}^{\mu}$, and a decoder $f(\boldsymbol{z}^{\mu})=f(g(\boldsymbol{x}^{\mu})):=\tilde{\boldsymbol{x}}^{\mu}$
that decompresses the latent feature representation into a reconstructed
output $\tilde{\boldsymbol{x}}^{\mu}$. The reconstruction quality
is commonly assessed using a loss function that measures the disparity
between the original input and the reconstructed output. We use the
mean-square error (MSE) between the input $\boldsymbol{x}^{\mu}$
and the output $\boldsymbol{\tilde{x}}^{\mu}$ as the loss function
for the autoencoder.

Overfitting arises when a model excessively fits the training data,
leading to worse performance on unseen data. To avoid overfitting,
we introduce an additional $L_{2}$ regularization into the loss function,
which discourages the model from acquiring overly intricate patterns
within the training data and encourages sparsity of the latent feature
output:

\begin{equation}
L=\frac{1}{P}\overset{P}{\underset{\mu=1}{\sum}}||\boldsymbol{x}^{\mu}-\tilde{\boldsymbol{x}}^{\mu}||^{2}+\gamma\underset{i}{\sum^{M}}||\boldsymbol{\theta}_{i}^{2}||,
\end{equation}

where $P$ represents the number of data in the dataset ($P=300$),
$\boldsymbol{x}^{\mu}$ and $\tilde{\boldsymbol{x}}^{\mu}$ represent
the original input and reconstructed output for the $\mu^{th}$ observation,
$\boldsymbol{\theta}_{i}$ represents the $i^{th}$ parameters (weights
and biases) in the functions $f$ and $g$. Using the sum of the squared
difference between the original input and the reconstructed output
for the $\mu^{th}$ observation means that the reconstructed outputs
are encouraged to be close to the original inputs. For regularization
terms added to mitigate the overfitting problem, $\gamma$ is a regularization
parameter that controls the trade-off between the reconstruction error
and the penalty term. As $\gamma$ approaches zero, the weight of
the $L_{2}$ regularization term becomes very small and has almost
no penalizing effect on the parameters. This makes the model focus
more on reducing the loss function itself and less on the sparsity
of the parameters. When $\gamma$ tends to infinity, the weight of
the $L_{2}$ regularization item becomes very large, which leads to
a stronger tendency to choose a sparse feature model. $\gamma$ can
be arbitrary, and it is usually a small number (as regularization
and reconstruction compete with each other, we typically pick $\gamma=10^{-6}$
to ensure good reconstruction quality).

\subsubsection{Convolutional Autoencoder \label{subsec:Convolutional-Autoencoder introduction}}

In the following, we outline the architecture of the autoencoder used
for such reconstruction tasks. Fig.\ref{fig:autoencoder}(a) illustrates
the general structure of an autoencoder, which consists of two main
parts: the encoder and the decoder. 

Encoder: The encoder starts with an input layer, which expects images
of shape $28\times28\times1$. It then applies a series of layers
to this input: (1) a convolutional layer with 32 filters, each of
size $3\times3$ (with linear activation); (2) a max pooling layer
with a pool size of $2\times2$; (3) a convolutional layer with 32
filters, each of size $3\times3$ (with ReLU activation); (4) a max
pooling layer with a pool size of $2\times2$; (5) a dense layer with
2 units (with ReLU activation). This layer represents the bottleneck
layer, which is the encoded version of the input.

Decoder: The decoder takes the output of the encoder and applies a
series of layers to reconstruct the original input: (1) a transposed
convolutional (deconvolutional) layer with 32 filters, each of size
$3\times3$ and a stride of 2 (with ReLU activation); (2) a transposed
convolutional layer with 32 filters, each of size $3\times3$ and
a stride of 2 (with ReLU activation); (3) a convolutional layer with
1 filter of size $3\times3$ (with linear activation). This layer
reconstructs the original input from the encoded representation. 

We use the same padding throughout all the convolutional layers.

\subsection{Binary classification\label{subsec:Binary-classification Introduction}}

\label{sec:Binary-classification}

For a given input sample $\boldsymbol{x}^{\mu}$, we use the sigmoid
function $\sigma$ as our classifier, and the predicted output $\hat{y}^{\mu}$
is calculated as follows:

\begin{equation}
\hat{y}^{\mu}=\sigma(\boldsymbol{w}^{T}\boldsymbol{x}^{\mu})=\frac{1}{1-e^{-\boldsymbol{w}^{T}\boldsymbol{x}^{\mu}}}.\label{eq:Activate Func}
\end{equation}

To generate the maximum-margin solution within our capacity, we use
cross-entropy as our loss function under the condition where the weight
will be the most robust and generalizable to other similar inputs

\begin{equation}
\mathcal{L}=\frac{1}{P}\underset{\mu}{\sum}\left[y^{\mu}\ln\hat{y}^{\mu}+(1-y^{\mu})\ln(1-\hat{y}^{\mu})\right],\label{eq:Loss Func}
\end{equation}

where $P$ refers to the number of data samples $\boldsymbol{x}^{\mu}$,
and $\mu$ indexes the data. $y^{\mu}$ and $\hat{y}^{\mu}$ refer
to the actual and predicted outputs for the $\mathord{\mu}^{th}$
input data $\boldsymbol{x}^{\mu}$, respectively. 

We use gradient descent $\frac{\partial\mathcal{L}}{\partial\overrightarrow{w}}$
based on the loss function $\mathcal{L}$ to update the weight, which
is given by

\begin{equation}
\frac{\partial\mathcal{L}}{\partial\overrightarrow{w}}=\frac{1}{p}\underset{\mu}{\sum}\left(\hat{y}^{\mu}-y^{\mu}\right)\boldsymbol{x}^{\mu}.\label{eq:Gradient descend}
\end{equation}

Our classification result is a binary distribution as a function of
$\hat{y}^{\mu}$. Given data $\boldsymbol{x}^{\mu}$, the probability
of the true label $y^{\mu}$ being equal to 1 is $\hat{y}^{\mu}$

\begin{equation}
P(y^{\mu}=1\mid\boldsymbol{x}^{\mu})=\hat{y}^{\mu},\label{eq:Probability of y_mu =00003D +1}
\end{equation}

and given data $\boldsymbol{x}^{\mu}$, the probability of true label
$y^{\mu}$ equals to $-1$ is $1-\hat{y}^{\mu}$

\begin{equation}
P(y^{\mu}=-1\mid\boldsymbol{x}^{\mu})=1-\hat{y}^{\mu}.\label{eq:Probability of y_mu =00003D -1}
\end{equation}

\subsubsection{Principal Component Analysis (PCA) \label{subsec:PCA-and-correlation Introduction}}

PCA is used to analyze large datasets with high dimensions, improving
data interpretability while retaining the maximum amount of information
and enabling multidimensional data display. It aims to find a projection
of the data onto directions that maximize the variance of the original
dataset. The PCA algorithm can reduce the dimensionality of data to
provide a clearer understanding and visualization. Additionally, the
PCA algorithm is fast and does not require parameter tuning or optimization.
However, its disadvantage is that it is based solely on covariance
($2^{nd}$ order statistics) and is limited to linear projections.
PCA can also be used to denoise and compress data, as well as identify
informative variables that better explain the data.

Given a dataset $\left\{ \boldsymbol{x}^{\mu}\right\} _{\mu=1}^{P}$,
we subtract the mean $\left\langle \boldsymbol{x}\right\rangle $
from the data, resulting in $\left\{ \boldsymbol{x}^{\mu}-\left\langle \boldsymbol{x}\right\rangle \right\} _{\mu=1}^{P}.$
Then, we find the principal components $\boldsymbol{e}_{i},i=1,...,N$
of the dataset. We then project the data onto the first $M$ principal
directions ($M=2$ in the main text) and add back the mean to obtain
the reconstructed data,
\begin{equation}
\tilde{\boldsymbol{x}}^{\mu}=\sum_{i=1}^{M}\left(\boldsymbol{x}^{\mu}-\left\langle \boldsymbol{x}\right\rangle \right)^{T}\boldsymbol{e}_{i}+\left\langle \boldsymbol{x}\right\rangle 
\end{equation}

\subsubsection{t-distributed Stochastic Neighbor Embedding (t-SNE)\label{subsec:t-SNE-embedding Introduction}}

t-SNE is a statistical method , an embedding model (data is usually
embedded in two or three dimensions) for reducing high-dimensional
data and visualizing data. Basically, it transfers the distance of
high-dimensional data points to probability distribution and models
a similar low-dimensional data using these distances. 

Assume we have a $N$ dimension data $\boldsymbol{x}_{1-N}$. Firstly,
the t-SNE algorithm constructs a probability distribution over high-dimensional
object pairs so that points that are dissimilar are allocated a lower
probability, while objects that are similar are assigned a greater
probability. Then, we generate the same amount of low-dimensional
data at random $y_{1-N}$.

For each high-dimensional data point $x_{i}$, we define a conditional
probability distribution $p_{j\mid i}$, where $i\neq j$

\begin{equation}
p_{j\mid i}=\frac{\exp(-\parallel x_{i}-x_{j}\parallel^{2}/2\sigma_{i}^{2})}{\sum_{k\neq i}\exp(-\parallel x_{i}-x_{k}\parallel^{2}/2\sigma_{i}^{2})},
\end{equation}

and set $p_{i\mid i}=0$. Note that $\underset{j}{\sum}p_{j\mid i}=1$
for all $i$. Then we define the similarity of high-dimensional data
as $p_{ij}$

\begin{equation}
p_{ij}=\frac{p_{j\mid i}+p_{i\mid j}}{2N}.
\end{equation}

For low-dimensional data $y_{1-N}$, we similarity define a $q_{ij}$
\begin{equation}
q_{ij}=\frac{(1+\parallel y_{j}-y_{i}\parallel^{2})^{-1}}{\sum_{k}\sum_{l\ne k}(1+\parallel y_{k}-y_{l}\parallel^{2})^{-1}}.
\end{equation}

We would like to measure the difference between two probability distributions
(high-dimensional distribution $p_{ij}$ and low-dimensional distribution
$q_{ij}$), with respect to the location of data points. Here we use
Kullback\textendash Leibler divergence (KL divergence) to calculate
the difference. We use KL divergence as the lost function $L$,

\begin{equation}
L=KL(p\parallel q)=\sum_{i,j}p_{i,j}\log\frac{p_{i,j}}{q_{i,j}}.
\end{equation}

Finally, we use gradient descent to minimize the KL divergence and
update low-dimensional data to make it as similar as the high-dimensional
data as possible until the algorithm converges. 

After these steps, the t-SNE algorithm maps data from a high-dimensional
space to a low-dimensional space, which forms clusters in low-dimensional
data and preserves a local atlas of high-dimensional data. \\

\subsubsection{Receiver Operating Characteristic (ROC) curve\label{subsec:ROC-curve Introduction}}

The ROC (Receiver Operating Characteristic) curve is a graphical representation
that showcases the diagnostic ability of a binary classifier system.
It is created by plotting the TPR (true positive rate) against the
FPR (false positive rate) at various classification thresholds. For
models with fixed thresholds, data that has been classified can result
in four different scenarios, namely $TP$ (True Positive), $TN$ (True
Negative), $FP$ (False Positive,Type I error), and $FN$ (False Negative,Type
II error).

The $TPR$ is the ratio of correctly classified positive instances
to all positive instances ($P$), which means the number of real positive
cases in the data:

\begin{equation}
TPR=\frac{TP}{P}=\frac{TP}{TP+FN}.
\end{equation}

The $FPR$ is the ratio of incorrectly classified negative instances
to all negative instances, which means the number of real negative
cases in the data:

\begin{equation}
FPR=\frac{FP}{N}=\frac{FP}{FP+TN}.
\end{equation}

The ROC curve provides a comprehensive picture of the classifier's
performance by illustrating the trade-off between the true positive
rate and the false positive rate. In an ideal scenario, the TPR is
1 and the FPR is 0, implying a perfect classifier that makes no false
predictions. This ideal situation is represented by a point on the
top-left corner of the ROC curve.

The area under the ROC curve (AUC) is used as a measure of the classifier's
performance. The AUC ranges from 0 to 1, with a higher value indicating
better discrimination ability. An AUC of 0.5 signifies that the classifier
performs no better than random guessing, while an AUC of 1 represents
a perfect classifier. On the contrary, when the value is 0, the result
is exactly the opposite of the answer.

The ROC curve provide a simple and intuitive way to compare and select
the best model among multiple candidates. A model with a higher AUC
is generally preferred as it demonstrates better predictive accuracy.
Moreover, it offer a visual representation of the classifier's quality
and help choose an optimal classification threshold based on the desired
trade-off between the true positive rate and the false positive rate. 

\bibliographystyle{unsrt}
\bibliography{xampl,refs}

\end{document}